\documentclass{elsart3p}

\usepackage{graphics}
 \usepackage{epsfig}

\usepackage{amssymb}
\newcommand{\beq}{\begin{equation}}
\newcommand{\eeq}{\end{equation}}
\newcommand{\qvec}{{\bf q}}

\begin{document}

\begin{frontmatter}
%titles, authors and addresses

\title{Low-energy signatures of charge and spin fluctuations in Raman and optical
spectra of the cuprates}


\author{S. Caprara, C. Di Castro, T. Enss, and M. Grilli}
%\author[a,b]{James Q. Thirdauthor}

\address{SMC - Istituto Nazionale per la Fisica della Materia, and 
Dipartimento di Fisica, Universit\`a ``Sapienza'', Roma, Italy}

\begin{abstract}
We calculate the optical and Raman response within a phenomenological model of 
fermion quasiparticles coupled to nearly critical collective modes. We
find that, whereas critical scaling properties might be masked in optical 
spectra due to charge conservation, distinct critical signatures of 
charge and spin fluctuations can be detected in Raman spectra exploiting
specific symmetry properties. We compare our results with recent experiments
on the cuprates.
\end{abstract}

\begin{keyword}
% keywords here, in the form: keyword; keyword
Stripes; Quantum critical point; spectroscopy

\PACS 74.72.-h \sep 78.30.-j \sep 71.45.Lr \sep 74.20.Mn
\end{keyword}
\end{frontmatter}

% main text
%\section{Title of the first section}
%\label{labelOfFirstSection}
%%%%%%%%%%%%%%%%%%%%%%%%%%%%%%%%%%  main text %%%%%%%%%%%%%%%%%%

\section{Introduction}
It has been proposed that the anomalous physical properties of the cuprates 
arise from the proximity to a quantum critical point (QCP) located near 
optimal doping \cite{review1,varma,tallon,CDG}. The onset of the ordered 
phase, however, can be prevented by the low dimensionality of the CuO$_2$ 
layers, by disorder, and/or by the occurrence of pairing, thereby making the 
order that should prevail in the low-temperature underdoped region rather 
elusive. Besides the possibility of unusual ordered states \cite{varma,walter}, 
experiments indicate that charge ordering (CO), possibly in the form of stripe 
or checkerboard textures, is indeed present in underdoped superconducting 
cuprates. A survey is provided in Refs. \cite{review1,review2}. 

Based on the experimental evidence and on theoretical results within models 
for strongly correlated electrons, a scenario was proposed \cite{CDG}, where 
quantum criticality near optimal doping and CO in the underdoped region  
merge in a unified scenario. Furthermore, due to the closeness to an antiferromagnetic 
(AF) Mott-insulating phase at very low doping, also the presence of 
substantial AF spin fluctuations could markedly affect the physical properties 
of these systems \cite{chubukov}. The roles of AF and charge fluctuations are 
not necessarily exclusive: The former could rather be emphasized by CO, and 
be ``enslaved'' in the charge-poor regions produced by the modulation of the 
charge profile, thus extending their relevance to doping 
levels substantially higher than the AF onset.

Within a model of strongly interacting electrons coupled to a dispersionless 
phonon \cite{CDG}, for realistic model parameters, the CO instability first 
occurs along the $(1,0)$ or $(0,1)$ directions (i.e., along the Cu-O bonds) 
with typical wavelength of a few lattice spacings. Accordingly, 
near the QCP the fermion quasiparticles (QP) are coupled to low-energy CO 
collective modes (CM), and are strongly affected near the ``hot'' spots at the 
Fermi surface, which are mutually connected by th critical wavevector
${\bf q}_c$ and reside near the 
points $(\pm\pi,0)$ and $(0,\pm\pi)$ of the Brillouin zone. A similar 
phenomenology is found in the case of AF spin fluctuations, although with a 
different characteristic wavevector, ${\bf q}_s\approx (\pi,\pi)$. The effects 
of the exchange of nearly critical CM have been extensively studied within 
phenomenological models with QP-CM coupling. In this way, specific features in 
ARPES (e.g., the QP self-energy \cite{sulpizi}, the kink \cite{SG1} and the 
isotopic dependence of the dispersions \cite{SG2}), and anomalous isotopic 
effects \cite{andergassen} have been coherently explained within the CO-CM 
scenario. 

In this paper, we explore the optical and Raman signatures of the CO and of 
the (enslaved) AF spin fluctuations, in particular (a) the possibility to 
detect spectroscopic features displaying scaling properties as fingerprints of 
quantum critical CM and (b) how the symmetry properties of the Raman selection 
rules determine the relevance of the spin and/or charge CM at different energy 
scales.

\section{Critical collective modes in optical spectra}
Near the CO instability (i.e., near the QCP at optimal doping and in the 
underdoped, on the verge of CO), the CO CM have a small 
mass and provide a ``cheap'' reservoir of excitations, which can easily affect 
the various spectroscopic probes.

Recently, we performed a {\it microscopic} calculation of the optical 
conductivity $\sigma(\omega)$ \cite{CGDE}, based on a conserving approximation 
of the current-current response, which is crucial to correctly describe 
optical absorption. It was important to realize that for a parabolic electron 
dispersion and in the absence of quenched impurities (clean case), the CM 
cannot be purely electronic and must involve, e.g., phonons, otherwise the 
conservation of electron momentum would entail a vanishing $\sigma(\omega)$ at 
$\omega>0$. Phonons modify the purely electronic relaxational form of the CO 
CM propagator, yielding an effective interaction
\beq 
D({\bf q}, \omega_n)=-\frac{g^2}{\nu ({\bf q}-{\bf q}_c)^2+|\omega_n|+
\omega_n^2/{\overline \Omega}+m} \label{propagator}
\eeq
among the QP. Here, $\omega_n$ are boson Matsubara frequencies, $g$ is the 
QP-CM coupling, $\nu$ is a fermion scale, the high-frequency cutoff 
$\overline\Omega$ relates fermion and phonon scales \cite{CGDE} and encodes 
the dynamical nature of the phonons, and $m\propto\xi^{-2}$ is the CO CM mass, 
proportional to the inverse square correlation length measuring the distance 
to the CO transition. A similar effective interaction describes also the AF 
CM, with a proper redefinition of the parameters, and yields very similar 
results for $\sigma(\omega)$. For ${\overline\Omega}<\infty$, a finite 
response was found, with various regimes for $\sigma(\omega)$ \cite{CGDE}. At 
low temperature $T$, the near absence of dissipation for 
$\omega<\overline\Omega$ strongly suppresses $\sigma(\omega)$, yielding a peak 
which is fixed at $\omega\sim\overline\Omega$ and does not scale with $m$ when 
$m<\overline\Omega$. On the other hand, if $m>\overline\Omega$, a peak related 
to the excitation of CO CM appears at $\omega\sim m$. At higher $T$, the 
thermal excitation of the diffusive CM rapidly fills the low-frequency 
spectrum, embedding the finite-frequency peaks into a broader peak at 
$\omega=0$. Therefore, the scaling behavior may be masked by non-universal 
effects, and we concluded that the lack of scaling peaks in optical spectra 
cannot be taken as evidence against quantum criticality.

%%%%%%%%%%%%%%%%%%%%%%%%%%%%%%%%%%%%%%%%%%%%%%%%%%%%%%%%%%%%%%%%%%%%%%
\begin{figure}
\includegraphics[scale=0.3]{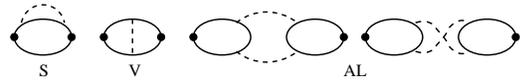}
%\vspace{-4truecm}
\caption{Diagrams for the CM corrections to the response function. The full 
dots represent the current vertices in the case of optical conductivity or the 
Raman vertices. The solid  lines represent the QP propagator (with or without 
impurity scattering) and the dashed lines represent the interaction mediated 
by CO or AF CM. Self-energy, vertex, and Aslamazov-Larkin-like contributions
are labeled by S, V, and AL, respectively.}
\label{fig.1}
%
%       fig. 1: 
\end{figure}
%%%%%%%%%%%%%%%%%%%%%%%%%%%%%%%%%%%%%%%%%%%%%%%%%%%%%%%%%%%%%%%%%%%%%%%%

Here, we address the question whether and how the presence of impurities 
(dirty case) modifies the above conclusions. We determine the conditions for a 
finite-frequency peak to be visible and display scaling  via its 
mass-dependent position, when an impurity scattering rate $\Gamma$ broadens 
the QP propagator in Fig. \ref{fig.1}. In these diagrams the impurity vertex corrections
cancel out, while, owing to the finite $\qvec_c$, 
no diffusive impurity ladders can be inserted.
We assume a CM mass $m\propto T$, as 
appropriate for the quantum-critical regime, and calculate the corrections 
$\delta\sigma(\omega)$ arising from the current-current diagrams of Fig. 
\ref{fig.1}. We first discuss the purely perturbative result, and then 
introduce a suitable resummation scheme to avoid the shortcomings of 
perturbation theory.

Impurities enlarge the free-electron $\delta(\omega)$ response into a Drude 
peak of width $\Gamma$. If $\Gamma$ is of the order of (or even larger than) 
$m$ and/or $\overline \Omega$, this Drude term embeds the peaks due to CM 
scattering and makes the critical behavior of the fluctuations difficult to 
detect. Therefore, a first conclusion is that the observation of peaks due to 
CM is only possible when disorder is sufficiently weak. To further explore the 
effects of critical CM in optics we accordingly introduce a rather small 
disorder scattering rate $\Gamma <m,\overline \Omega$.  Disorder produces a 
finite negative part in the CM perturbative correction, 
$\delta\sigma(\omega)$, at low $\omega$. This is naturally so, because the 
conservation of the optical weight implies that $\delta\sigma$ has zero total 
weight, $\int d\omega \delta\sigma(\omega)=0$. Therefore, when the dirty QP 
are dressed by CO CM, the spectral weight added at frequencies 
$\omega\gtrsim\Gamma$ has to be subtracted at $\omega\lesssim\Gamma$. This 
feature was already present in the clean case, where, however, the negative 
part was $\delta(\omega)$-like and did not affect the spectra at finite 
$\omega$ \cite{CGDE}. 

\begin{figure}
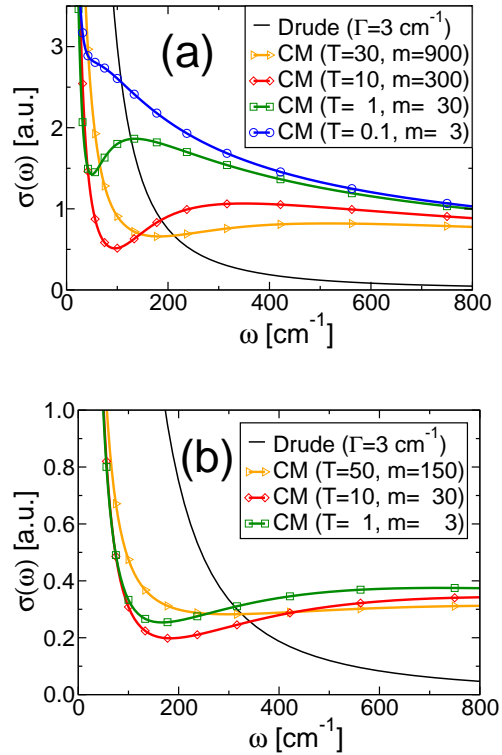

{\includegraphics[scale=0.25]{sigscal.eps}\vspace{0.65truecm}}\\
\includegraphics[scale=0.25]{sigfiss.eps}
\caption{Optical conductivity for a system of QP coupled to CO CM modes in the 
presence of finite impurity scattering rate $\Gamma=3$ cm$^{-1}$, at various 
temperatures and CM masses, for (a) $\overline \Omega= 30$ cm$^{-1}$, and (b) 
$\overline \Omega=330$ cm$^{-1}$.}
\label{fig.2}
%
%       fig. 2: 
\end{figure}

%
%As in the clean case, when $m>\overline\Omega$, the 
%excess of spectral weight is peaked at $\omega\sim m\propto T$, reflecting the 
%quantum critical behavior of the CM. On the other hand, the peak is pinned at 
%$\omega\sim\overline\Omega$ when $m<\overline\Omega$. Again, this lack of 
%scaling spectral features is a consequence of charge conservation and occurs 
%despite the fully quantum-critical behavior of the CO CM ($m\propto T$).
%
%We emphasize that a perturbative scheme cannot provide a reliable description 
%of the interplay of CM and impurity scattering at low frequency. 
%
In the presence of impurities, the 
negative part of $\delta\sigma$ may prevail despite the Drude peak even for 
small or moderate QP-CM coupling $g$, yielding an unphysical negative 
absorption at low $\omega$. This artifact can be cured by an appropriate 
resummation scheme, which we propose here. Specifically, we consider a memory 
function \cite{goetze}, which treats on equal footing the QP scattering due to 
impurities and to nearly critical CM, 
$M(\omega)=i\Gamma -\omega \chi_{CM}^{clean}(\omega)/W$, where 
$\chi_{CM}^{clean}$ is the {\it clean} CM correction to the current-current 
response (see Fig. \ref{fig.1}) and $W$ is the free QP optical weight. The 
corresponding resummed response is
\beq
\chi_{jj}(\omega)=\frac{W \omega}{\omega+M(\omega)},
\label{memorychi}
\eeq
which reproduces the perturbative result at high $\omega$, where 
$\chi_{CM}^{clean}/W\ll 1$. In the absence of QP-CM coupling 
$\chi_{CM}^{clean}=0$ and we recover the bare Drude response 
$\chi_D=W\omega/(\omega+i\Gamma)$. Although our resummation scheme is no 
longer conserving, it does not spoil the previous conclusions on the 
conditions for optical peaks scaling with $m$. Figs. \ref{fig.2}(a) and (b) 
report the results for the resummed 
$\sigma(\omega)=-{\rm Im}\chi_{jj}(\omega)/\omega$ for the cases 
$m>\overline\Omega$ and $m<\overline\Omega$, respectively.

%%%%%%%%%%%%%%%%%%%%%%%%%%%%%%%%%%%%%%%%%%%%%%%%%%%%%%%%%%%%%%%%%%%%%%%
%%%%%%%%%%%%%%%%%%%%%%%%%%%%%%%%%%%%%%%%%%%%%%%%%%%%%%%%%%%%%%%%%%%%%%%%

Our analysis shows that, in the presence of a small amount of impurities, 
the spectra broaden, but the generic feature persists, and the absence of a 
$T$-dependence in the peak position [see Fig. \ref{fig.2}(b)] cannot be used 
to infer the absence of critical CM (with quantum-critical mass $m\propto T$). 
The CM become fully visible in optical spectra only at low $T$ and when the CM 
mass is larger than a typical non-universal dissipation scale 
$\overline\Omega$ [see Fig. \ref{fig.2}(a)]. Note that the displacement of 
weight from $\omega\lesssim\Gamma$ to higher frequencies, produced by 
$\delta\sigma$, can result in a narrowing of the Drude peak. In the analysis 
of experiments, this could lead to an erroneous overestimate of the QP 
lifetime. The determination of the precise amount of narrowing and of the 
extent to which the negative part of $\delta\sigma$ could ``carve'' a dip in 
the spectrum, thereby producing a finite-frequency peak despite the rapid 
thermal filling mentioned above, requires a reliable conserving resummation 
scheme to treat the low-frequency absorption, which is presently not 
available, due to the lack of a small expansion parameter.
 
In summary, within our approach a sufficiently strong disorder may effectively 
wash out the effects of CM, thereby hiding criticality. When disorder is weak 
enough to leave the peaks at low $T$ visible, impurities do not qualitatively 
modify the findings of the clean case: The key role in the non-universal 
behavior of $\sigma(\omega)$ is played by charge conservation in the presence 
of an extrinsic (e.g., of phonon origin) energy scale for dissipation 
$\overline\Omega$.  The value of the critical scale $m$ in comparison to 
$\overline \Omega$ then determines the presence or the absence of a critical 
$T$ dependence of the optical features. In cuprates we estimate $\overline \Omega \sim
100$ cm$^{-1}$\cite{CGDE}.

\section{Critical collective modes in Raman spectra}
The natural question arises on the role of dissipation on the behavior of 
other spectral quantities, which are not ``protected'' by charge conservation. 
Within this respect, Raman spectroscopy provides a nice playground with the 
additional advantage of a momentum-dependent filtering of the spectra. A key 
difference with respect to the optical conductivity is that, properly 
adjusting the polarization of the incoming and outgoing photons, the Raman 
response involves form factors selecting specific regions of the QP Brillouin 
zone \cite{reviewraman}. This helps to distinguish nearly critical AF and CO 
fluctuations in Raman spectra, which would look very similar in optical 
spectra \cite{ECDG}.

\begin{figure}
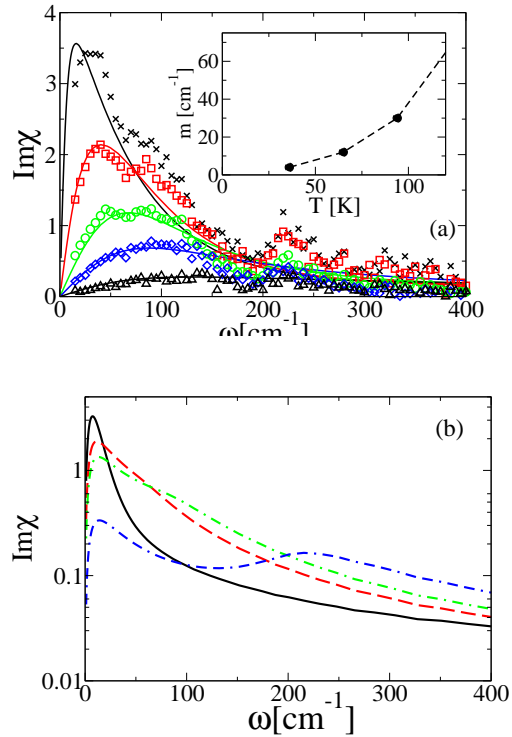

{\includegraphics[scale=0.25]{Fig3a-miami.eps}\vspace{0.6truecm}}\\
\includegraphics[scale=0.25]{Fig3b-miami.eps}
\caption{(a) Raman absorption in the $B_{1g}$ channel for a clean system of QP 
coupled to CO CM at temperatures (from top to bottom) $T$=36,65,94,124,188 
K. The mass $m(T)$ needed to fit the data is reported in the inset. A 
dissipation scale $\overline\Omega= 200$ cm$^{-1}$ is used. (b) Same as in (a) 
with $T$=36,94,124,188 K, but with a smaller dissipation scale, 
$\overline\Omega= 20$ cm$^{-1}$.}
%\vspace{-0.4truecm}
\label{fig.3}
%
%       fig. 3: 
\end{figure}

To compare Raman and optical spectra, we consider here the contribution to the 
Raman response due to AF and CO CM. Since we have shown that impurity 
scattering should be weak enough not to hide the effects of CM in optics, we 
henceforth focus on the role of CM and neglect impurities. Specifically, we 
consider the clean Raman response ${\rm Im}\chi$ in the $B_{1g}$ and $B_{2g}$ 
channels, where photons couple to QP via the vertices 
$\gamma_{B_{1g}}=\cos k_x-\cos k_y$ and $\gamma_{B_{2g}}=\sin k_x\sin k_y$. 
While $\gamma_{B_{1g}}$ is large (and therefore selects excitations) in the 
antinodal regions around the $(\pm\pi,0)$ and $(0,\pm\pi)$ points of the 
Brillouin zone and changes sign in the nodal regions $k_x=\pm k_y$, 
$\gamma_{B_{2g}}$ is large around the nodes and changes sign in the antinodal 
regions. These symmetry properties induce leading-order cancellations between 
the Raman response diagrams depending on the momentum exchanged by the CM. For 
instance, we find that the S and V diagrams (see Fig. 
\ref{fig.1}) cancel each other in the $B_{1g}$ channel, when typical momenta 
$\qvec_c$ of CO excitations are exchanged, because $\gamma_{B_{1g}}$ has the 
same value at the two hot spots connected by $\qvec_c$, and the S and V 
diagrams have intrinsically opposite sign\footnote{This cancellation of S and 
V diagrams is reminiscent of the loop cancellation occurring in the 
density-density response of systems with strong forward scattering 
\cite{MCD}.}. In this case, the only relevant contribution arises from the AL 
diagrams, which were considered in Ref. \cite{CDGS}, to explain the
Raman absorption anomalies observed in La$_{2-x}$Sr$_x$CuO$_4$ (LSCO) with 
$x=0.02,0.10$ \cite{tassini}. On the 
other hand, AF CM with typical wavevector ${\bf q}_s\approx(\pi,\pi)$ give 
qualitatively wrong results. In Ref. \cite{CDGS}, a diffusive CM propagator 
was considered, mediating an effective interaction similar to that in Eq. 
(\ref{propagator}), but with a different high-frequency behavior. Here, having 
in mind a comparison with the results on $\sigma(\omega)$, obtained using Eq. 
(\ref{propagator}), we reproduce the analysis of Ref. \cite{CDGS} using the 
high-frequency cutoff $\overline\Omega$. As it can be seen in Fig. 
\ref{fig.3}(a), the experimental data are well reproduced setting 
$\overline\Omega=200$ cm$^{-1}$ and using $m$ as the only fitting parameter, 
once the overall intensity is set by the curve at the highest temperature. 
Again, $m(T)$ behaves as expected for a system crossing over from the quantum 
critical region to the low-$T$ nearly ordered phase. Clearly, due to the 
inherently non-conserving character of Raman excitations, the scaling behavior 
of the peak is visible also when $m<\overline \Omega$. For the sake of 
completeness, we also discuss the case $m >\overline\Omega$, although it does
not apply to LSCO. In Fig. \ref{fig.3}(b) we report the $B_{1g}$ Raman spectra 
(again coming from the AL diagrams only, for symmetry reasons) for the same 
masses as in Fig. \ref{fig.3}(a), but a smaller $\overline \Omega =20$ 
cm$^{-1}$. Here, a narrow peak appears at a frequency 
$\omega\sim\overline\Omega$, above which the CM changes from diffusive to 
propagating [see Eq. (\ref{propagator})]. The mass energy scale $m$ is still 
visible as a shoulder (for low and intermediate $T$) or as a broad peak (at 
the largest temperature $T=188$ K).

%%%%%%%%%%%%%%%%%%%%%%%%%%%%%%%%%%%%%%%%%%%%%%%%%%%%%%%%%%%%%%%%%%%%%%%
%%%%%%%%%%%%%%%%%%%%%%%%%%%%%%%%%%%%%%%%%%%%%%%%%%%%%%%%%%%%%%%%%%%%%%%%

\section{Symmetry properties of  Raman spectra}
As mentioned in the previous section, the specific symmetry of Raman vertices 
was exploited in Ref. \cite{CDGS} to explain the anomalies observed in LSCO 
\cite{tassini}. In this section, we focus on those features of Raman spectra 
which can be discussed based only on symmetry considerations. These provide 
rather robust arguments on the relevance or irrelevance of the 
{\it leading} contributions from CO or AF CM in the various channels when 
the CM are nearly critical and strongly peaked at their characteristic 
wavevectors. On the other hand, far away from criticality, when $m$ is large 
and the CM mediate smooth and weak interactions, the symmetry arguments are of 
little use because the CM correction to the response is small and substantial 
contributions come from regions of the Fermi surface away from the hot spots. 
 
By direct inspection of the symmetry properties of the QP loops entering the AL 
diagrams (see Fig. \ref{fig.1}), one realizes that the leading contributions 
of AF CM with ${\bf q}_s=(\pi,\pi)$ are non-zero only in the 
$\gamma_{A_{1g}}=\cos(k_x)+\cos(k_y)$ symmetry \cite{venturini}. In the $B_{1g}$ 
channel, a leading contribution to Raman absorption from the AL diagrams can 
only arise if the CO CM have a characteristic wavevector $\qvec_c$ in the 
$(1,0)$ or $(0,1)$ directions \cite{CDGS}. As far as the S and V diagrams are 
concerned, as anticipated in the previous section, the leading contributions 
in the $B_{1g}$ symmetry cancel when the CO CM is considered. In the $B_{2g}$ 
symmetry, instead, the S and V diagrams sum up (since now the Raman vertex has 
opposite sign in the two hot spots connected by ${\bf q}_c$, avoiding loop
cancellation) and provide a finite leading contribution. The opposite 
situation is found when the CM wavevector is $\qvec_s\approx(\pi,\pi)$, 
typical of AF CM near the AF instability: due to the symmetry of the form 
factors, the S and V diagrams sum up in the $B_{1g}$ channel, while they 
cancel at leading order in the $B_{2g}$ channel. The situation is summarized 
in Table 1.

%%%%%%%%%%%%%%%%%%%%%%%%%%%%%%%%%%%%%%%%%%%%%%%%%%%%%%%%%%%%%%%%%%%%%%%%%%%
%%%%%%%%%%%%%%%%%%%%%%%%%%%%%%%%%%%%%%%%%%%%%%%%%%%%%%%%%%%%%%%%%%%%%%%%%%%
\begin{table}
%\begin{center}
\begin{tabular}{|c|c|c|}
\hline
& $B_{1g}$ &  $B_{2g}$ \\
\hline
CO CM & $AL\ne 0$, \, $S+V\approx 0$  &  $AL\approx 0$, \, $S+V\ne 0$  \\
\hline
AF CM & $AL\approx 0$, \, $S+V\ne 0$  & $AL\approx 0$, \, $S+V\approx 0$   \\
\hline
\end{tabular}
%\end{center}
\caption{Leading corrections to the Raman absorption due to the 
processes represented by the diagrams of Fig. \ref{fig.1}, in the $B_{1g}$
and  $B_{2g}$ symmetries.}
\end{table}   
%%%%%%%%%%%%%%%%%%%%%%%%%%%%%%%%%%%%%%%%%%%%%%%%%%%%%%%%%%%%%%%%%%%%%%%%%%%
%%%%%%%%%%%%%%%%%%%%%%%%%%%%%%%%%%%%%%%%%%%%%%%%%%%%%%%%%%%%%%%%%%%%%%%%%%%

\begin{figure}
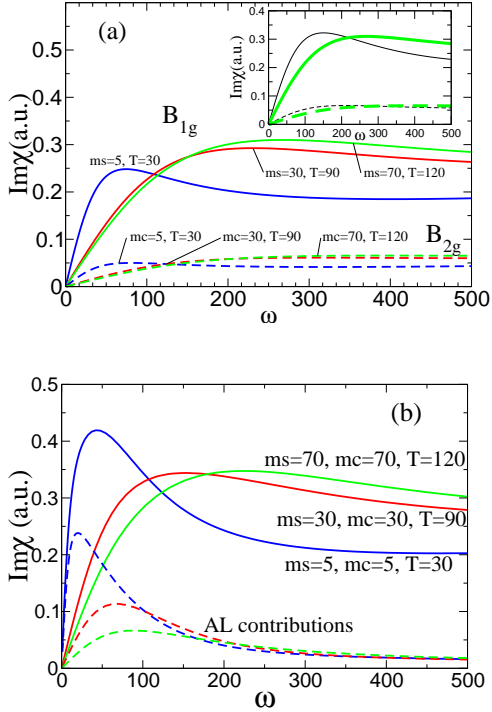

{\includegraphics[scale=0.25]{SVSCr.eps}\vspace{0.65truecm}}\\
\includegraphics[scale=0.25]{TOT.eps}
\caption{(a) Raman absorption for a clean system of QP due to S+V processes. 
Solid lines: $B_{1g}$ channel with QP coupled to AF CM with 
$\overline\Omega_{AF}=500$ cm$^{-1}$; dashed line: $B_{2g}$ channel with QP 
coupled to CO CM with $\overline\Omega=200$ cm$^{-1}$. AF and CO CM have 
masses $ms$ and $mc$, respectively. Inset: Same as in the main panel, but with 
fixed $T=120$ cm$^{-1}$ and two different masses of the CM: $m=70$ cm$^{-1}$ 
(thick lines) and $m=200$ cm$^{-1}$ (thin lines). (b) Total $B_{1g}$ Raman 
absorpion due to AF CM (S+V diagrams) and to CO CM (AL diagrams, also shown 
separately as dashed lines). Masses and temperatures are the same as in panel 
(a).}
%\vspace{-0.4truecm}
\label{fig.4}
%
%       fig. 4: 
\end{figure}

Since the leading S+V contributions diverge as $\omega\to 0$, a resummation is 
required to obtain a finite Raman response. We adopt the same memory-function 
scheme of Eq. (\ref{memorychi}) (here we consider the clean case $\Gamma=0$). 
We calculate the corrections to Raman absorption due to both AF and CO CM in 
the two channels. Specifically, in the inset of  Fig. \ref{fig.4}(a) we report 
the changes in the $B_{1g}$ and $B_{2g}$ Raman absorption due to a progressive 
reduction of the CM mass, keeping $T$ and QP-CM coupling fixed. This situation 
is representative of experiments comparing overdoped samples (large CM mass) 
with optimally and underdoped samples (with progressively smaller CM mass). 
The main effect is a shift of weight from low to high $\omega$. The same shift 
occurs upon reducing both $T$ and CM masses, as can be seen in the main panel 
of Fig. \ref{fig.4}(a), representative of the quantum critical regime for the 
CM. Notice that, according to the standard expression of Raman vertices 
\cite{reviewraman}, in this figure the intensity of the $B_{2g}$ spectra has 
been rescaled by a factor $t'/t\sim 0.2$, where $t'$ and $t$ are the next 
nearest neighbor and the nearest neighbor hopping parameters of a 
tight-binding model, respectively. Moreover, while the $B_{2g}$ channel never 
acquires leading AL corrections from AF or CO diagrams (see Table 1), the 
$B_{1g}$ absorption may be enhanced if CO CM become increasingly critical when 
the doping is reduced. In this case, their additional AL contribution appears 
at frequencies $\omega\lesssim\overline\Omega\sim 200\dots 300$ cm$^{-1}$. In 
Fig. \ref{fig.4}(b) we report the $B_{1g}$ absorption arising from the 
superposition of AF CM acting via S+V processes (resummed within a 
memory-function scheme to remove their singular behavior at low frequency) and 
CO CM in the AL channel. Notice that here CO and AF CM have the same mass to
schematize their enslaved quantum critical behavior.

%%%%%%%%%%%%%%%%%%%%%%%%%%%%%%%%%%%%%%%%%%%%%%%%%%%%%%%%%%%%%%%%%%%%%%%
%%%%%%%%%%%%%%%%%%%%%%%%%%%%%%%%%%%%%%%%%%%%%%%%%%%%%%%%%%%%%%%%%%%%%%%%
%%%%%%%%%%%%%%%%%%%%%%%%%%%%%%%%%%%%%%%%%%%%%%%%%%%%%%%%%%%%%%%%%%%%%%%%%%%

Several of the above features are indeed generically observed in LSCO and 
Bi$_2$SrCa$_2$Cu$_2$O$_8$ (BSCCO) \cite{ruditom}. In the $B_{1g}$ channel, upon 
going from overdoped to underdoped systems, weight is shifted to higher 
frequencies. According to our scheme, this effect could be attributed to 
AF CM acting via S+V processes, and becomes progressively more 
important upon underdoping, both in LSCO and BSCCO. The effect of CO CM via 
S+V processes is nearly cancelled in the $B_{1g}$ channel and the anomalous absorption,
which we interpret as due to the AL 
contribution, is rather weak even around optimal doping. Upon entering the 
underdoped region, the marked tendency of LSCO to form stripes makes the AL 
contribution of CO to the $B_{1g}$ channel more important at low frequencies 
($\omega\lesssim 200$ cm$^{-1}$). Then, while the $B_{1g}$ absorption stays 
moderately suppressed by the AF S+V contributions at intermediate frequencies, 
it acquires a nearly critical additional contribution from the AL processes 
\cite{CDGS} and displays the anomalous low-frequency absorption shown in Fig. 
\ref{fig.4}(b). Notice that a similar additional anomalous absorption is 
absent in BSCCO samples. We argue that in BSCCO samples the tendency to form 
nearly static stripes is much weaker, so that the low-frequency AL absorption 
could be masked by other processes competing with CO (e.g., pairing without 
superconducting phase coherence). Another possibility (not necessarily 
exclusive) is that in BSCCO the CO texture has the different form of a 
checkerboard (previously named ``eggbox'' in the literature \cite{goetzEPJ}). 
In this case, symmetry arguments can be found showing that checkerboard charge 
fluctuations should couple much less to the QP loop entering the AL diagrams 
of Fig. \ref{fig.1} even for the $B_{1g}$ symmetry. Of course these 
non-trivial signatures naturally require more specific parameter tuning and 
more realistic model calculations (e.g., with specific band structures) and 
treatment of the cutoffs to improve the quantitative agreement with the 
behavior of various samples. 

In conclusion, the robust symmetry arguments outlined above account for the 
generic behavior of the Raman spectra at different dopings and in different 
channels and, contrary to the optical conductivity, allow to distinguish the 
effects of charge and spin modes.

{\bf Acknowledgments} We acknowledge interesting discussions with C. 
Castellani, T. Devereaux, R. Hackl, and J. Lorenzana and financial support from 
the MIUR-PRIN 2005 - prot.\ 2005022492 and (CDC and TE) from the Alexander von 
Humboldt foundation.

%\appendix

%\section{Title of the first section of the appendix}
%\label{labelFirstSecOfAppendix}
%Insert your text of the first section of the appendix here.

\end{document}